\documentclass[a4paper,12pt]{article}%
\usepackage{amsfonts}
\usepackage{amsmath}
\usepackage{amssymb}
\usepackage{graphicx}%
\providecommand{\U}[1]{\protect\rule{.1in}{.1in}}
\begin{document}

\title{Mass Formula for Light Nonstrange Mesons and Regge\ Trajectories in Quark Model}
\author{Duojie JIA$^{1}\thanks{Corresponding author: E-mail: jiadj@nwnu.edu.cn}$, Cheng-Qun Pang$^2$, A. Hosaka$^3$ \\$^{1}$\textit{Institute of Theoretical Physics, College of Physics and}\\\textit{Electronic Engineering, Northwest Normal University,\ }\\\textit{Lanzhou 730070, China}\\$^{2}$\textit{College of Physics and Electronic Information, Qinghai }\\\textit{Normal University,Xining China}\\$^{3}$\textit{Research Center for Nuclear Physics (RCNP), Osaka }\\\textit{University, Ibaraki, Osaka 567-0047, Japan}}
\maketitle

\begin{abstract}
We study the Regge-like spectra of light mesons in a relativized quark model.
An analytical mass formula is presented for the light unflavored mesons with
the help of auxiliary field method, by which a quasi-linear
Regge-Chew-Frautschi plot is predicted for the orbitally excited states. We
show that the trajectory slope is proportional to the inverse confining
parameter $1/a$ up to a factor depending on the strong coupling $\alpha_{s}$
when the orbital quantum number $L$ is large. The result is tested against the experimental data of the spectra of the meson families $\pi/b$,$\rho/a$%
,$\eta/h$ and $\omega/f$ in the ($L$,$M^{2}$) planes, with the fitted
parameters consistent with that in the literatures.

{\normalsize PACS number(s):14.40.Be,12.40.Nn, 12.39.Ki}

{\normalsize Key Words Light-light mesons, Regge spectra, Quark model}

\end{abstract}

\section{Introduction}

It is remarkable experimentally that most hadrons consisting of light quarks
fall on straight lines known as "Regge trajectories" \cite{ChewF:61},
\begin{equation}
M^{2}=(1/\alpha^{\prime})(J-\alpha_{0}). \label{JM}%
\end{equation}
This simple relation, referred to as Regge-Chew-Frautschi plot, enable us to
group the hadrons with mass $M$ and the angular momentum $J$ into a series of
rotational families'. The observation of linear behavior (\ref{JM}) in the
hadron spectrum dates back to the 1970's\cite{Collons:77} and remains to be
the subject of recent discussions as new states are discovered. In the case of
light nonstrange mesons (which we shall discuss in this work) composed of
unflavored quark and antiquark the slope parameter $\alpha^{\prime}$ varies
only slightly from family to family (by less than $10\%$)
\cite{AnisovichAS:PRD00,MasjuanAB:D12,Sharov:13,Selem:06,SonnenscheinW:JHEP14}%
, by which the linearity of trajectories are assumed. The validity of the
linear Regge trajectories were addressed in the literatures, for instance, in
\cite{InopinS:2001,TangN:PRD00,GuoWW:D08} on the experimental ground. The
correction to the linear trajectory are explored in
\cite{JohnsonN:PRD79,BrisudovaBG:00,Selem:06}, and recently in Ref.
\cite{SonnenscheinW:JHEP14} for the light to heavy mesons systematically.

In the picture of AdS/QCD, Forkel et al\cite{Forkel:07a,Forkel:07b} predicted
the mass relation to be of the Regge-like%
\begin{equation}
M^{2}=4\lambda^{2}(J+1/2), \label{MJ}%
\end{equation}
for the ground state of the mesons. In quark model, however, the linear
behavior of the Regge trajectories remains to be
understood\cite{KangS:75,MaungKN:D93,EbertFG:D09,SilvestreSB:JPA2008}(to name
a few). For further discussions on the Regge-like mesonic spectra, see Refs.
\cite{Afonin:A07,Afonin:C07,KlemptZ:P07,ShifmanV:D08,Afonin:JA07,Afonin:JA08}, for instance. For recent discussions, see Afonin\cite{AfoninP14,AfoninPA14,Affonin:16} and Masjuan at al.\cite{MasjuanAB14} and references therein.

Purpose of this work is to revisit the Regge-like spectra of the light
unflavored mesons in relativized quark model. With the help of auxiliary field
(AF) method, an analytical mass formula is derived for the orbitally excited
unflavored mesons, by which a quasi-linear Regge-Chew-Frautschi plot is
predicted. We find that the trajectory slope is inversely proportional to the
confining parameter $a$ up to a factor depending on the (averaged) strong
coupling $\alpha_{s}$ and the orbital quantum number $L$. We also test the
obtained mass formula by fitting the recent observed data of the light meson
families', $\pi/b$,$\rho/a$,$\eta/h$ and $\omega/f$ in the ($L$,$M^{2}$)
planes. The best fitted values of model parameters($a=0.164GeV^{2}$, the
vacuum constant $V_{0}=-343$MeV and $\alpha_{s}=0.57$) are given and in
consistent with that of relativized quark models in the literatures. The
limitation of the mass formula is discussed.

The $\eta/h$ trajectory is found to be distinctive (the best fit gives
$a=0.226GeV^{2}$ and $V_{0}=-751$MeV), and it suggests that the members of
this trajectory have the obvious $\bar{s}s$ component, for which the
zero-quark-mass approximation in the model is invalid.

\section{Quark model for light unflavored mesons}

We use the relativized quark model with Hamiltonian given by
\cite{Durand:D82,LichtenbergNP:pl82,Durand:D84,GodfreyIsgur:85,Jacobs:D86},
with the spin-dependent interactions suppressed for simplicity. The model
Hamiltonian is%

\begin{equation}
H=\sum_{i=1}^{2}\sqrt{\mathbf{p}^{2}+m_{i}^{2}}+V, \label{SH}%
\end{equation}
where
\begin{equation}
V=ar-\frac{4\alpha_{s}}{3r}+V_{0}, \label{V}%
\end{equation}
\ In (\ref{V}), the first term is the linear confining potential, with the
confining parameter $a$, the second is the color-Coulomb potential with
$\alpha_{s}$ the strong coupling constant, and $V_{0}$ the vacuum constant.
Here, $r=|\mathbf{x}_{1}-\mathbf{x}_{2}|$ is the relative coordinate of the
quark $1$ and antiquark $2$, with the bare masses $m_{1}$ and $m_{2}$,
respectively. We assume in our analysis that the two quarks are equal in mass,
and make no attempt to differentiate between them. $V$ is the effective
confining potential with the form of the linear plus Color-Coulomb parts. In
the sector of heavy flavor this fom of the interquark potential is confirmed
by Lattice QCD\cite{Bali01}. For the recent interquark potential in Lattice
simulation, see \cite{KawanaiS:12}, with $a=$ $\,0.155(19)$ reported.

We are mainly interested in the quantum-mechanical spectrum of the light
mesons in the analytical form. To this aim, the auxiliary field (AF)
method\cite{AMPolyakov:87,GubankovaYD:PLB94,MorgunovNS:PLB99} is employed to
transform the Hamiltonian into an analytically solvable one. The idea of the
AF method is to apply the equality $\sqrt{B}=\min_{\lambda}\{\frac{B}%
{2\lambda}+\frac{\lambda}{2}\}$ (the minimization is achieved when
$\lambda=\sqrt{B}$, $\lambda$ is positive) to rewrite the model Hamiltonian
(\ref{SH}) as $H=\min_{\mu_{1,2},\nu}\left\{  H(\mu_{1,2},\nu)\right\}  $,
where
\begin{equation}%
\begin{array}
[c]{c}%
H(\mu_{1,2},\nu)=\sum_{j=1}^{2}\left[  \frac{\mathbf{p}^{2}+m_{j}^{2}}%
{2\mu_{j}}+\frac{\mu_{j}}{2}\right]  +\frac{a^{2}r^{2}}{2\nu}+\frac{\nu}{2}\\
-\frac{4}{3}\frac{\alpha_{s}}{r}+V_{0},
\end{array}
\label{Hmn}%
\end{equation}
and the auxiliary fields, denoted as $\mu_{j}$($j=1,2$) and $\nu$ here, are
operators in the quantum-mechanical sense. As a result, $H(\mu_{j},\nu)$ is
equivalent to (\ref{SH}) up to the elimination of the fields $(\mu_{1},\mu
_{2},\nu)$ with the help of the constraints%
\begin{equation}%
\begin{array}
[c]{c}%
\delta_{\mu_{j}}H(\mu_{1,2},\nu)=0\Longrightarrow\mu_{j}\rightarrow\mu
_{j,0}=\sqrt{\mathbf{p}_{j}^{2}+m_{j}^{2}},\\
\delta_{\nu}H(\mu_{1,2},\nu)=0\Longrightarrow\nu\rightarrow\nu_{0}%
=a|\mathbf{x}_{1}-\mathbf{x}_{2}|=ar.
\end{array}
\label{min}%
\end{equation}
Here, the expectation average $\langle\mu_{i,0}\rangle$ ($i=1,2$) can be
viewed as the dynamical mass of the quark $i$, and $\langle\nu_{i,0}\rangle$
as the static energy of the flux-tube (QCD string) linking the quark $1$ and
$2$\cite{SemaySN:2004,SilvestreSB:JPA2008}.

Although the auxiliary fields are operators quantum-mechanically, the
calculations simplify considerably if one considers them as real
$c$-numbers\cite{SilvestreSB:JPA2008}. They are to be eliminated, in the AF
method, eventually through a minimization of the mass spectrum with respect to
the AF's, for which the optimal values of $\mu_{1,2}$ and $\nu$ are logically
close to $\langle\mu_{i,0}\rangle$ and $\langle\nu_{i,0}\rangle$,
respectively. For more details of the AF method applied to the mesons, see
\cite{KalashnikovaN:PLB2000,SemaySN:2004,SilvestreSB:JPA2008,Simonov:PL88,KalashnikovaN:PAN97}, and the references therein.

In the center of mass systems where the total momentum vanishes, the
Hamiltonian (\ref{Hmn}) becomes
\begin{equation}%
\begin{array}
[c]{c}%
H(\mu_{1,2},\nu)=\frac{\mathbf{p}^{2}}{2\mu}+\frac{1}{2}\mu\omega^{2}%
r^{2}+\frac{\mu_{M}+\nu}{2}\\
-\frac{A}{2ar}+\frac{m_{1}^{2}}{2\mu_{1}}+\frac{m_{2}^{2}}{2\mu_{2}}+V_{0}%
\end{array}
\label{HCM}%
\end{equation}
in which $\mu_{M}=\mu_{1}+\mu_{2}$, $\mu=\mu_{1}\mu_{2}/\mu_{M}$ is the
reduced mass of quark system, and

\begin{equation}
\omega=\frac{a}{\sqrt{\mu\nu}}, \label{Omg}%
\end{equation}%
\begin{equation}
A=\frac{8}{3}\alpha_{s}a. \label{A}%
\end{equation}

Since the first line of (\ref{HCM}) is simply the harmonic oscillator
Hamiltonian, one can, of course, diagonalize the Hamiltonian $H(\mu_{1,2}%
,\nu)$ in (\ref{HCM}) in the harmonic oscillator basis $|nLm\rangle$. As a
result, the quantized energy $E_{N}(\mu_{1,2},\nu)=\langle|H(\mu_{1,2}%
,\nu)|\rangle_{nLm}$ of the Hamiltonian (\ref{HCM}) is

\begin{equation}%
\begin{array}
[c]{c}%
E_{N}(\mu_{1,2},\nu)=\omega\left(N+\frac{3}{2}\right)  -\frac{A}{2\nu}\\
+\frac{\mu_{M}+\nu}{2}+\frac{m_{1}^{2}}{2\mu_{1}}+\frac{m_{2}^{2}}{2\mu_{2}%
}+V_{0},
\end{array}
\label{Emn}%
\end{equation}
where $N=n+L$, with $n$ and $L$ the radial quantum number and the orbital angular momentum of the bound system, respectively.
In deriving (\ref{Emn}), we have estimated the contribution of the
color-Coulomb term by its quantum average:

\begin{equation}%
\begin{array}
[c]{c}%
-\frac{4}{3}\left\langle \frac{\alpha_{s}}{r}\right\rangle \approx
-\frac{4\alpha_{s}}{3\langle|\mathbf{x}_{1}-\mathbf{x}_{2}|\rangle}\\
=-\frac{A}{2\nu},
\end{array}
\label{DEc}%
\end{equation}
where in the last step a simple relation $\langle|\mathbf{x}_{1}%
-\mathbf{x}_{2}|\rangle=\nu/a$, obtained from (\ref{min}), has been used. In addition, we use the notation $\alpha_{s}$ again to stand for the expectation $\langle\alpha_{s}\rangle$ for simplicity. But one should bear in mind that $\alpha_{s}$ hereafter represents an appropriate average of the QCD running coupling which differs in implication from the same symbol in (\ref{V}).

We note that we write presumably the band quantum number $N$ of the harmonic oscillator in (\ref{Emn}) in the form $N=n+L$, rather than $N=2n+L$, considering that the superfluous symmetry enters in the Hamiltonian (\ref{Hmn}) which is absent in the original Hamiltonian (\ref{SH}) when the AF method applied: $r\rightarrow r^{2}$, and it may bring some unphysical degeneracy in the radially excited states that may not be there in the spectrum of the model (\ref{SH}). Another reason is that the ratio of the slope parameters for the radial and angular-momentum trajectories is approximately 1:1, as observed by Anisovich et al.\cite{AnisovichAS:PRD00} and suggested by Afonin \cite{Afonin:A07,Afonin:C07,AfoninP14,AfoninPA14}, Klempt \cite {KlemptZ:P07} as well as Shifman et al. \cite{ShifmanV:D08}. For the orbitally excited states with $n=0$, which we shall consider in this work, it is expected that the superfluous symmetry does not show up.

\section{Mass formula and Regge trajectory}

We consider only, in this work, the orbitally excited mesons for which the
radial quantum number $n$ is set to be zero. To solve the model (\ref{SH})
with the AF method, one has to minimize the energy (\ref{Emn}) and thereby
eliminate the three auxiliary fields appearing in (\ref{Emn}). This amounts to
solving simultaneously the three constraints
\[
\partial_{I}E_{N}(\mu_{1,2},\nu)=0(I=\mu_{1},\mu_{2},\nu),
\]
which are explicitly
\begin{align}
\frac{a_{N}\nu}{2\sqrt{(\mu\nu)^{3}}}\left(  \frac{\mu_{2}}{\mu_{M}}-\frac
{\mu_{1}\mu_{2}}{\mu_{M}^{2}}\right)   &  =\frac{1}{2}-\frac{m_{1}^{2}}%
{2\mu_{1}^{2}},\label{cons1}\\
\frac{a_{N}\nu}{2\sqrt{(\mu\nu)^{3}}}\left(  \frac{\mu_{1}}{\mu_{M}}-\frac
{\mu_{1}\mu_{2}}{\mu_{M}^{2}}\right)   &  =\frac{1}{2}-\frac{m_{2}^{2}}%
{2\mu_{2}^{2}},\label{cons2}\\
\frac{a_{N}\mu}{\sqrt{(\mu\nu)^{3}}}  &  =1+\frac{A}{\nu^{2}}, \label{cons3}%
\end{align}
where we denote $a_{N}\equiv a(L+3/2)$.

Since $m_{i}$ is same for quark and antiquark, one has, by symmetry, $\mu_{1}%
=\mu_{2}=2\mu$. It follows that $\mu_{i}=\mu_{M}/2$($i=1,2$), or equivalently,
$\mu=\mu_{0}\equiv\mu_{M}/4$. Hence, (\ref{cons1}) and (\ref{cons2}) simplify%

\begin{equation}
\frac{a_{N}\nu}{4(\mu\nu)^{3/2}}=1-\frac{4m^{2}}{\mu_{M}^{2}},. \label{EqB}%
\end{equation}
By putting (\ref{EqB}) into (\ref{cons3}), one finds, with a little algebra%

\begin{equation}
\mu_{M}=\frac{4a_{N}^{2}\nu}{(\nu^{2}+A)^{2}}=4\mu_{0}, \label{muM}%
\end{equation}
or,%
\begin{equation}
\mu_{1}=\mu_{2}=2\mu_{0}=\frac{2a_{N}^{2}\nu}{(\nu^{2}+A)^{2}}. \label{mu12}%
\end{equation}

Having $\mu$ as a function of $\nu$ one can solve $\nu$ by rewriting the
energy (\ref{Emn}) as an energy functional of $\nu$. We assume the bare mass
$m\simeq0$ as it should be quite small at the scale of meson mass. When using
(\ref{mu12}), the energy (\ref{Emn}) becomes,%

\begin{equation}
E_{N}(\mu_{0},\nu)=\frac{3}{2}\nu+\frac{A}{2\nu}+\frac{2a_{N}^{2}\nu}{(\nu
^{2}+A)^{2}}+V_{0}, \label{Ev}%
\end{equation}
which is minimized by the constraint equation (that is, $0=\delta_{\nu}%
E_{N}(\mu_{0},\nu)$)%

\begin{equation}
3-\frac{A}{\nu^{2}}+\frac{4a_{N}^{2}}{(\nu^{2}+A)^{2}}-\frac{16a_{N}^{2}%
\nu^{2}}{(\nu^{2}+A)^{3}}=0. \label{AF}%
\end{equation}

To solve (\ref{AF}), we firstly consider the case $A/\nu^{2}\ll1$. To the
lowest order, $A/\nu^{2}\rightarrow0$ yields
\begin{equation}
\nu^{2}=2a_{N}. \label{v2a}%
\end{equation}
By making ansatz $\nu^{2}=2a_{N}+c_{A}A$ and putting it into (\ref{AF}), one
can show
\[
c_{A}=-\frac{3}{2},
\]
which leads to, to the first order of $A/\nu^{2}$,%
\begin{equation}
\nu^{2}\simeq2a_{N}-\frac{3A}{2}. \label{v22}%
\end{equation}

Hence, the AF equations (\ref{cons1}) through (\ref{cons3}) can be solved by
(\ref{mu12}) as well as
\begin{equation}
\nu_{0}=\sqrt{2a\left(  L+\frac{3}{2}-2\alpha_{s}\right)  }. \label{Nuu}%
\end{equation}
Therefore, one finds, from (\ref{Ev})
\begin{equation}
\langle H\rangle_{N}=\left(  \frac{3}{2}+\frac{1}{2c_{N}}\right)  \nu
_{0}+\frac{A}{2\nu_{0}}+V_{0} \label{Hv0}%
\end{equation}
with $\nu_{0}$ given by (\ref{Nuu}). In the case of $\mu_{1}=\mu_{2}$, one
obtains a mass formula for the unflavored $\bar{q}q$ mesons,
\begin{equation}
M_{\bar{q}q}=\left(  \frac{3}{2}+\frac{1}{2c_{N}}\right)  \sqrt{2a\left(
L+\frac{3}{2}-2\alpha_{s}\right)  }+\frac{4a\alpha_{s}}{3\sqrt{2a\left(
L+\frac{3}{2}-2\alpha_{s}\right)  }}+V_{0}, \label{Mqq}%
\end{equation}
in which
\begin{align}
c_{N}  &  =\frac{(\nu^{2}+A)^{2}}{4a_{N}^{2}}\nonumber\\
&  =\left(  1-\frac{2\alpha_{s}}{3(L+3/2)}\right)  ^{2}. \label{cN}%
\end{align}
We note that $c_{N}\rightarrow1$ when $L$ becomes large.

It follows, by squaring (\ref{Mqq}), that
\begin{equation}
\left(  M_{\bar{q}q}-V_{0}\right)^{2}=2aw_{N}^{2}\left[  L+\frac{3}%
{2}-2\alpha_{s}+\frac{4}{3}\frac{\alpha_{s}}{w_{N}}+\frac{(4\alpha_{s}%
/3)^{2}}{(2L+3-4\alpha_{s})w_{N}^{2}}\right]  , \label{QRegg}%
\end{equation}
where $w_{N}\equiv\left(  3/2+1/(2c_{N})\right)  $ tends to $2$ when $L$ is
large. When one ignores the last term in the brackets in the RHS of
(\ref{QRegg}), it leads to a quasi-linear from of the Regge-Chew-Frautschi
plot
\begin{equation}
\left(  M_{\bar{q}q}-V_{0}\right)  ^{2}=2a\left(  \frac{3}{2}+\frac{1}{2c_{N}%
}\right)  ^{2}\left[  L+\frac{3}{2}-\frac{5\alpha_{s}}{3}\right]  .
\label{QRegge}%
\end{equation}
Here, the limit $w_{N}\rightarrow2$ has been applied at which $c_{N}%
\rightarrow1$.

We see that this quasi-linear plot is comparable to the linear Regge
trajectories (\ref{JM}), if $V_{0}$ is small compared to the meson scale, that
is, $V_{0}/M_{\bar{q}q}\ll1$. Further, one can see that the slope parameter
$\alpha^{\prime}$ of the trajectory (\ref{QRegge}) depends upon $L$ weakly,
\begin{equation}
\alpha^{\prime}=\frac{1}{2a}\left(  \frac{3}{2}+\frac{1}{2}\left(
1-\frac{2\alpha_{s}}{3(L+3/2)}\right)  ^{-2}\right)  ^{-2}, \label{Slope}%
\end{equation}
in which the relation (\ref{cN}) is used. This slope increases slowly with the
quantum number $L$, and tends to $(8a)^{-1}$ when $L$ is very large:
$\lim_{L\rightarrow\infty}$($\alpha^{\prime})=(8a)^{-1}$. The slope
$(8a)^{-1}$ was also derived in Ref.\cite{KangS:75} using the WKB
approximation. It is to be compared to the slope $1/(2\pi a)$ predicted by the
relativistic string model \cite{Goto:71,GoddardGR:73,Nambu:74}.

The strong coupling $\alpha_{s}$ enters also in both the slope (\ref{Slope})
and the intercept,
\begin{align}
-\alpha_{0}& =\frac{3}{2}-2\alpha_{s}+\frac{4}{3}\frac{\alpha_{s}}{w_{N}%
}+\frac{(4\alpha_{s}/3)^{2}}{(2L+3-4\alpha_{s})w_{N}^{2}},\label{al0}\\
&  \simeq\frac{3}{2}-\frac{5\alpha_{s}}{3}, \label{all}%
\end{align}
with%
\[
w_{N}\equiv\frac{1}{2}\left[  3+\left(  1-\frac{4\alpha_{s}}{3(2L+3)}\right)
^{-2}\right]  .
\]

It is remarkable that while the intercept (\ref{al0}) depends on the strong
coupling $\alpha_{s}$ and $L$ it does not depend on the confining parameter
$a$. The trajectory slope (\ref{Slope}) is inversely proportional to $a$ when
$L$ is large. We note that after the treatment of the color-Coulomb
interaction using the AF method in section 2, $\alpha_{s}$ is in fact the
averaged values of the strong coupling $\alpha_{s}(r)$ depending on the
interquark distance $r$: $\alpha_{s}\rightarrow\langle|\alpha_{s}|\rangle_{N}%
$, and therefore we do not expect it to agree quantitatively with that of the
values measured by QCD lattice simulations.

Although (\ref{Mqq}), or (\ref{QRegg}), is quite suited to actual test against
the observed spectrum of light mesons, we shall conclude this section with
discussion of the extreme situation $A/\nu^{2}\sim1$ which may go beyond
perturbative regime for solving (\ref{v2a}) from the AF equation (\ref{AF}).

In the perturbative treatment from (\ref{v2a}) to (\ref{v22}), we assumed
$A/\nu^{2}\ll1$, which may be violated when $L$ is small, say, $L=0$. This
assumption can not be justified by only requiring $A/(2a_{N})$ to be small,
which is case ( $A/(2a_{N})<0.54$) even in the worst case $L=0$ for the
parameter setup of the Godfrey-Isgur(GI) model: $a=0.18GeV^{2}$, $\alpha
_{s}\simeq0.6$. At this typical setup, one can show numerically from
(\ref{AF}) that $A/\nu^{2}$ decreases from $0.69$ to $0.13$, while in the case
of (\ref{Nuu}) it drops rapidly from $2.67$ to $0.15$, as shown in Table 1
and Table 2, respectively.

One sees that the perturbative solution (\ref{Nuu}) is less stable near $L=0$
than the numerical one, and fails to satisfies $A/\nu^{2}\ll1$. We expect,
however, that this defect can be cured by the nice
linearity\cite{AnisovichAS:PRD00} of the trajectories of light mesons, as seen
in the linear fit(the solid line) shown in \textrm{FIG. 1}. When fitting
(\ref{Mqq}) to the data, we choose the tendency of the theoretical slope in
the ($L,M^{2}$) plot to agree with that of linear fitting without the first
($L=0$) states in the trajectories, that is, without the states $\rho
(770)1^{--}$, $\eta(548)0^{-+}$,$\omega(782)1^{--}$, respectively. In the case
of the $\pi/b$-trajectory, we retain the $L=1$ state $b_{1}(1235)1^{+-}$ since
the $L=0$ state in this trajectory is actually the pion, which we exclude in
this work due to its abnormally low mass.

\vspace{3mm}

{\small TABLE 1. The estimate for }$A/\nu^{2}$ {\small in the case of the
parameter setup of the GI model}, {\small made by the numerical solution(}%
$\nu_{0}${\small ) to (\ref{Hv0})}. {\small The resulted trajectory parameters
defined by (\ref{QRegg}) are also listed.}

\begin{tabular}
[c]{ccccccc}\hline\hline
$L$ & $0$ & $1$ & $2$ & $3$ & $4$ & $5$\\\hline
$A/\nu^{2}$ & $0.69$ & $0.37$ & $0.26$ & $0.19$ & $0.16$ & $0.13$\\
$\alpha^{\prime}(GeV^{-2})$ & $0.86$ & $0.80$ & $0.77$ & $0.75$ & $0.74$ &
$0.74$\\
Intercept($\alpha_{0}$) & $1.12$ & $0.65$ & $0.58$ & $0.56$ & $0.55$ &
$0.54$\\\hline\hline
\end{tabular}

\vspace{3mm}
{\small TABLE 2. The estimate for }${A/\nu}^{2}$ {\small for the parameter
setup of the GI model}, {\small made by solution (\ref{Nuu}), with the
corresponding trajectory parameters listed also, defined by (\ref{QRegg}).}

\begin{tabular}
[c]{ccccccc}\hline\hline
$L$ & $0$ & $1$ & $2$ & $3$ & $4$ & $5$\\\hline
$A/\nu^{2}$ & $2.67$ & $0.62$ & $0.35$ & $0.24$ & $0.18$ & $0.15$\\
$\alpha^{\prime}(GeV^{-2})$ & $0.86$ & $0.80$ & $0.77$ & $0.75$ & $0.74$ &
$0.74$\\
Intercept($\alpha_{0}$) & $1.12$ & $0.65$ & $0.58$ & $0.56$ & $0.55$ &
$0.54$\\\hline\hline
\end{tabular}
\vspace{3mm}

From \textrm{Table 1} one sees that $A/\nu^{2}\ll1$ is qualitatively justified
when $L$ $\geq2$. Generally, the smaller $\alpha_{s}$, the easier for this
requirement to hold. In addition, the trajectory slope $\alpha^{\prime}$
increases slowly with $L$, while the intercept $\alpha_{0}$ decreases slowly
when $L\geq2$. To reduce effects due to the derivation of $\nu$ from the
perturbative solution (\ref{Nuu}) at small $L$, we combine the guiding from
the linear trajectory with the parameter range predicted by the relativized
quark models
\cite{Durand:D82,LichtenbergNP:pl82,Durand:D84,GodfreyIsgur:85,Jacobs:D86}
when confronting (\ref{Mqq}) with the data. Though the GI setup of the model
parameters may not be the best fit\footnote{It may still be the best setup in
the sense that it reproduces whole mesonic spectra globally. We use, in this
work, the Godfrey-Isgur parameter setup as the initial setup for the search
regime during fitting.} in order to reproduce the trajectories of the four
families' considered, it is quite useful for the parameter searching for the
best fit, as illustrated in the following section.

\begin{figure}
[ptb]
\begin{center}
\includegraphics[
natheight=6.295000in,
natwidth=8.970900in,
height=3.3719in,
width=4.7937in
]%
{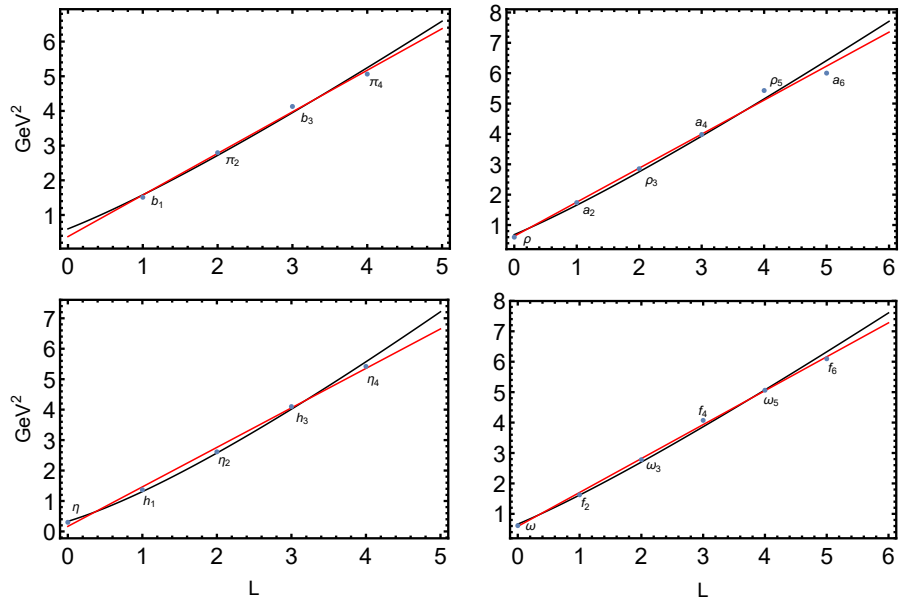}%
\caption{The linear (the solid lines) and the quasi-linear (the dashed lines)
fit of the masses of the meson members in the trajectoires of the $\pi
/b,\rho/a,\eta/h$ and $\omega/a$ families. The quasi-linear fit is done by the
mass formula (\ref{Mqq}). The solid circles correspond to the observed data.}%
\end{center}
\end{figure}

\section{The parameters and masses}

In this work, the light mesons are all assumed to be pure $\bar{q}q$ states
consisting of the up or down quarks with equal mass $m\simeq0$. To test the
mass formula (\ref{Mqq}), we choose four families of light mesons, marked by
$\pi/b$, $\rho/a$, $\eta/h$ and $\omega/f$, respectively. The corresponding
experimental data for the family members is taken entirely from the Particle
Data Group's (PDG) 2016 Review of Particle Physics\cite{Patrignani:2016}. Part
of the family members have been explored in Refs.
\cite{JohnsonN:PRD79,AnisovichAS:PRD00,Selem:06}, and a detailed studies were
given in Refs. \cite{EbertFG:D09,MasjuanAB:D12,SonnenscheinW:JHEP14}
associated with the Regge trajectory. The members of each trajectory has at
least three states. Meanwhile, they have a feature associated with $J^{PC}$
quantum number that the orbital quantum number $L$ changes successively in
trajectory, given that the pure $\bar{q}q$ state has the parity $P=(-)^{L+1}$
and $C=(-)^{L+S}$. To apply the linear-trajectory guiding mentioned in section
3 at relatively larger $L$ to the our quasi-linear fit, we list below some
details of the linear fits for the trajectories selected.

(i) The $\pi/b$ trajectory($I=1$): It includes the states $b_{1}%
(1235)1^{+-},\pi_{2}(1670$) $2^{-+}$, $b_{3}(2030)3^{+-}$, $\pi_{4}%
(2250)4^{-+}$. The lowest($L=0$) state, the pion, which should have been in
this trajectory, was excluded from trajectory analysis since it has abnormally
low mass. The corresponding linear fit($L=1,2,3,4$), to the most recently
observed data, is
\[
M^{2}(\pi/b)=1.199(L+0.314),\chi_{MS}^{2}=0.0205
\]
with the mean squared (MS) error $\chi_{MS}^{2}$ for the fit added. The fit is
depicted in the \textrm{FIG. 1}(a)(the solid line), including the experimental
data (the solid circles) for comparison. Here, the MS error is defined by
$\chi_{MS}^{2}=\sum_{L}(M_{L}^{Th}-M_{L}^{Exp})^{2}/(L_{\max}-1)$ where the
index $L$ runs from $0(1$ for the $\pi/b$ trajectory$)$ to the maximal value
$L_{\max}$ of the orbital number.

(ii) The $\rho/a$ trajectory($I=1$): The members chosen are the states
$\rho(770)1^{--}$, $a_{2}(1320)2^{++}$,$\rho_{3}(1690)3^{--}$, $a_{4}%
(2040)4^{++}$, $\rho_{5}(2350)5^{--}$ and $a_{6}(2450$) $6^{++}$. The linear
fit ($L=0,1,2,3,4,5$) and its MS error are
\[
M^{2}(\rho/a)=1.120(L+\allowbreak0.565),\chi_{MS}^{2}=0.0387
\]
The results of the fit is shown in the \textrm{FIG. 1}(b)(the solid line),
compared to the data(the solid circles). In the following the same mark
convention will be used when plotting the trajectories of the $\eta/h$ and
$\omega/f$).

(iii) The $\eta/h$ trajectory($I=0$): The members are the states
$\eta(548)0^{-+}$, $h_{1}(1170)1^{+-}$,$\eta_{2}(1645)2^{-+}$, $h_{3}%
(2025)3^{+-}$ and $\eta_{4}(2330)4^{-+}$. The result of the linear fit
($L=0,1,2,3,4$) is
\[
M^{2}(\eta/h)=1.297(L+0.128),\chi_{MS}^{2}=0.0181
\]
and depicted in the \textrm{FIG. 2 }(c)(the solid line).

(iv) The $\omega/f$ trajectory($I=0$): It is composed of the state members
$\omega(782)1^{--}$, $f_{2}(1270)2^{++}$, $\omega_{3}(1670)3^{--}$,
$f_{4}(2050)4^{++}$, $\omega_{5}(2250)5^{--}$ and $f_{6}(2510)6^{++}$. The
linear fit ($L=0,1,2,3,4,5$) gives
\[
M^{2}(\omega/f)=1.115(L+0.527),\chi_{MS}^{2}=0.0081
\]
and is depicted in the \textrm{FIG. 2 }(d)(the solid line).

One sees that in its n\"{a}tive form the linear formula (\ref{MJ}) applies for
the $\rho/a$ and $\omega/f$ trajectories for which $-\alpha_{0}\simeq0.5$, but
fails somehow in the case of the $\pi/b$ and $\eta/h$ trajectories for which
$-\alpha_{0}\simeq0.3$ and $0.1$, respectively.

We use (\ref{Mqq}), with $c_{N}$ given by (\ref{cN}), to fit the above four
meson families in the ($L$, $M^{2}$) plane (we set $n=0$). The optimal
parameters ($a,\alpha_{s}$,$V_{0}$) fitted to the data are shown in
\textrm{Table 2}, in which the notation Num. stands for the number of the data
coordinates. In \textrm{Table 3}, we show the mass values calculated from the
mass formula (\ref{Mqq}) and masses of the experiment (PDG)
\cite{Patrignani:2016}, with the corresponding MS error $\chi_{MS}^{2}$ listed also.

\vspace{3mm}
{\small TABLE 3. The optimal values for the confining parameter(}$a${\small ),
the strong coupling(}$\alpha_{s}$) {\small averaged and the vacuum constant
}$V_{0}$ {\small for fitting (\ref{Mqq}) to the experimental (PDG) mass
data\cite{Patrignani:2016}. The calculated trajectory parameters, when }%
$L=4${\small (for }$\pi/b,\eta/h${\small ) and }$L=5${\small (for }%
$\rho/a,\omega/f${\small ), and the MS error }$\chi_{MS}^{2}$ {\small for
fitting are also listed. }%

\begin{tabular}
[c]{cccccc}\hline\hline
Traj. & Num. & $a$(GeV$^{2}$) & $\alpha_{s}$ & $V_{0}$(GeV) & $\chi_{MS}^{2}%
$(GeV)\\\hline
$\pi/b$ & $4$ & $0.165$ & $0.584$ & $-0.351$ & $0.00299$\\
$\rho/a$ & $6$ & $0.167$ & $0.562$ & $-0.360$ & $0.00255$\\
$\eta/h$ & $5$ & $0.226$ & $0.581$ & $-0.751$ & $0.00335$\\
$\omega/f$ & $6$ & $0.159$ & $0.562$ & $-0.325$ & $0.00176$\\\hline\hline
\end{tabular}
$\ \ \ \ $
\vspace{3mm}

It can been seen from Table 3 that the optimal values of the model parameters
(confining parameter $a$, strong coupling $\alpha_{s}$, vacuum constant
$V_{0}$) fall in the regime around the GI setup, except for the $\eta/h$
trajectory for which $a$ and $-V_{0}$ are obviously larger. One can explain
this exceptional case as an indication that the members of $\eta/h$ trajectory
may have hidden $\bar{s}s$ component, which goes beyond the approximation
$m\simeq0$ taken in this work. When the $\eta/h$ trajectory removed, the
fitted values of $a,\alpha_{s}$ and $V_{0}$ vary slightly at the level of
$1.5\%$, $2.3\%$ and $9\%$ respectively. In this sense, we conclude that three
model parameters $a,\alpha_{s}$ and $V_{0}$ are universal, and has a globally
fit at about
\begin{equation}
a=0.164\text{GeV}^{2}\text{,}V_{0}=-343\text{MeV, }\alpha_{s}=0.57\text{.}
\label{best}%
\end{equation}

The trajectory slope at the global fit is
\begin{equation}
\alpha^{\prime}=0.871(6)\text{GeV}^{-2}\text{.} \label{alp}%
\end{equation}

For the intercept, no universal fit can be found. The above prediction for
$\alpha^{\prime}$ is in good agreement with that of the models in the
literatures, say, $0.886GeV^{2}$ in Ref.\cite{JohnsonN:PRD79}, $0.887GeV^{2}$
in Ref.\cite{EbertFG:D09}(for the $\rho$ parent trajectory), $0.884$ in
Ref.\cite{SonnenscheinW:JHEP14}, and $0.893$($=1/1.12$) in
Ref.\cite{MasjuanAB:D12}.

On the other hand, the study\cite{VeseliO:B96} of the parameters of the
relativized quark model provides a constraint on them: $200$MeV$\leq-V_{0}%
\leq470$MeV. This constraints are fulfilled by all fitted parameters in Table
3, except for the $\eta/h$ trajectory. Given that $\alpha_{s}$ is the averaged
value of the strong coupling in (\ref{SH}) the fit (\ref{best}) agrees well
with the parameter setup of the GI model\cite{GodfreyIsgur:85}(where
$\alpha_{s}\leq0.60$). For the parameter $a$, the fit in (\ref{best}) is
compatible with the results of the literatures:$a=0.180$ in
Ref.\cite{Durand:D84};$a=0.192$ in Ref.\cite{Jacobs:D86};$a=0.191$ in
Ref.\cite{FulcherCY:D93};$a=0.183$ in Ref.\cite{HwangK:D96}. We stress here
that our fit for $a$ in (\ref{best}) is closer to the confining parameter
$\sigma$ in the recent lattice simulation\cite{KawanaiS:12}: $(0.394(17))^{2}%
\simeq0.155(19)$.

\vspace{3mm}
{\small TABLE 4. The calculated masses(MeV) and that of the experiment (PDG)
\cite{Patrignani:2016} for the four light meson families. The states marked
with solid triangles are the established states given in the PDG summary
tables, while those marked an `f.' are the less established mesons given in
further states. The notation "Num." stands for the numerical prediction by solving the model (\ref{Mqq}).}
\begin{center}

\begin{tabular*}{14.3cm}
[c]{cccccc}\hline\hline
{\small Traj.} & $%
\begin{array}
[c]{r}%
\text{{\ mesons }}{\ \ }\\
\text{(}J^{PC}\text{ {\small Status})}%
\end{array}
$ & $\text{{\small Exp.}}$ & {\small This work} & {\small Num.} & $%

\begin{array}
[c]{r}%
\text{{\small Calculations in Refs.\ }}\\
\text{{\small GI\cite{GodfreyIsgur:85},EF\cite{EbertFG:D09}%
,SW\cite{SonnenscheinW:JHEP14}}}%
\end{array}
$\\\hline
$%
\begin{array}
[c]{r}%
\pi/b\ \ \\
(I=1)
\end{array}
$ & $%
\begin{array}
[c]{rrr}%
b_{1} & 1^{+-} & \blacktriangle\\
\pi_{2} & 2^{-+} & \blacktriangle\\
b_{3} & 3^{+-} & \text{f }\\
\pi_{4} & 4^{-+} & \text{f }%
\end{array}
$ & $%
\begin{array}
[c]{r}%
1229\\
1672\\
2030\\
2250
\end{array}
$ & $%
\begin{array}
[c]{r}%
1302\\
1666\\
1972\\
2242
\end{array}
$ & $%
\begin{array}
[c]{r}%
1241\\
1631\\
1951\\
2230
\end{array}
$ & $%
\begin{array}
[c]{rrr}%
1220 & 1258 & 1257\\
1680 & 1643 & 1650\\
2030 & 2164 & 1965\\
2330 & 2344 & 2236
\end{array}
$\\\hline
$%
\begin{array}
[c]{r}%
\rho/a\ \\
(I=1)
\end{array}
$ & $%
\begin{array}
[c]{rrr}%
\rho\  & 1^{--} & \blacktriangle\\
a_{2} & 2^{++} & \blacktriangle\\
\rho_{3} & 3^{--} & \blacktriangle\\
a_{4} & 4^{++} & \blacktriangle\\
\rho_{5} & 5^{--} & \\
a_{6} & 6^{++} &
\end{array}
$ & $%
\begin{array}
[c]{r}%
775\\
1318\\
1689\\
1995\\
2330\\
2450
\end{array}
$ & $%
\begin{array}
[c]{r}%
839\\
1310\\
1675\\
1982\\
2253\\
2497
\end{array}
$ & $%
\begin{array}
[c]{r}%
678.5\\
1243\\
1634\\
1956\\
2236\\
2487
\end{array}
$ & $%
\begin{array}
[c]{rrr}%
770 & 776 & 776\\
1310 & 1317 & 1324\\
1680 & 1714 & 1701\\
2010 & 2018 & 2008\\
2300 & 2264 & 2274\\
& 2475 & 2511
\end{array}
$\\\hline
$%
\begin{array}
[c]{r}%
\eta/h\ \ \\
(I=0)
\end{array}
$ & $%
\begin{array}
[c]{rrr}%
\eta & 0^{-+} & \blacktriangle\\
h_{1} & 1^{+-} & \blacktriangle\\
\eta_{2} & 2^{-+} & \blacktriangle\\
h_{3} & 3^{+-} & \text{f }\\
\eta_{4} & 4^{-+} & \text{f }%
\end{array}
$ & $%
\begin{array}
[c]{r}%
548\\
1170\\
1617\\
2025\\
2328
\end{array}
$ & $%
\begin{array}
[c]{r}%
638\\
1185\\
1611\\
1969\\
2285
\end{array}
$ & $%
\begin{array}
[c]{r}%
462.6\\
1124\\
1581\\
1955\\
2282
\end{array}
$ & $%
\begin{array}
[c]{rrr}%
520 &  & 545\\
1220 & 1485 & 1206\\
1680 & 1909 & 1612\\
2030 & 2209 & 1933\\
2330 & 2806 & 2208
\end{array}
$\\\hline
$%
\begin{array}
[c]{r}%
\omega/f\ \ \\
(I=0)
\end{array}
$ & $%
\begin{array}
[c]{rrr}%
\omega & 1^{--} & \blacktriangle\\
f_{2} & 2^{++} & \blacktriangle\\
\omega_{3} & 3^{--} & \blacktriangle\\
f_{4} & 4^{++} & \blacktriangle\\
\omega_{5} & 5^{--} & \text{f }\\
f_{6} & 6^{++} & \text{f }%
\end{array}
$ & $%
\begin{array}
[c]{r}%
783\\
1275\\
1667\\
2018\\
2250\\
2469
\end{array}
$ & $%
\begin{array}
[c]{r}%
846\\
1305\\
1661\\
1961\\
2225\\
2464
\end{array}
$ & $%
\begin{array}
[c]{r}%
695.4\\
1247\\
1628\\
1942\\
2215\\
2461
\end{array}
$ & $%
\begin{array}
[c]{rrr}%
780 &  & 768\\
1280 & 1529 & 1319\\
1680 & 1950 & 1698\\
2010 & 2286 & 2006\\
2300 & 2559 & 2271\\
& 2465 & 2509
\end{array}
$\\\hline\hline
\end{tabular*}
\end{center}
\vspace{3mm}

From Tables 4 we observe that the calculated masses from (\ref{Mqq}) are
systematically larger than the experimental values as well as that of the
references cited in the case of the low-$L$ states($L=0,1$). The ratio between
the prediction in this work and the experiment is about $1.04-1.14$ for the
($L=1$) states $b_{1}(1235)1^{+-}$ and the ($L=0$) states,$\rho(770)1^{--}$,
$\eta(548)0^{-+}$, $\omega(782)1^{--}$. This should not be surprise
considering that the perturbative solution (\ref{Nuu}) fails to satisfy
$A/\nu^{2}\ll1$ near $L=0$ and $1$, as seen from Table 1. This entails the
nonperturbatively solving of the AF equation (\ref{AF}) for the static energy
$\nu$ of string. One simple and direct way to do this is, for instance, to
promote (\ref{Nuu}) into an ansatz $\nu_{0}^{2}=2a\left(  L+\frac{3}%
{2}-f(\alpha_{s})\right)  $ and solve the unknown function $f(\alpha_{s})$
numerically or analytically (such a work is under way).

One of other limitations of the mass formula (\ref{Mqq}) comes from the chiral
limt (zero-mass limit) approximation for the light quark. This approximation
fails when meson has mixed hidden $\bar{s}s$(or $\bar{c}c$) component in its
internal structure. In this case, the calculation by the mixed state $|\bar
{q}q\rangle+|\bar{s}s\rangle$ is needed, for which the strange quark mass
$m_{s}$ is expected to enter in the mass formula (\ref{Mqq}). Another
limitation comes from the two-quark state ($\bar{q}q$) assumption for mesons.
This assumption may not be true when a meson has exotic structure, such as
components of gluoballs and/or multiquark states. This goes beyond the picture
of the native quark model used in this work and may make the mass formula
(\ref{Mqq}) insufficient.

\section{Summary}

We re-visit the orbital Regge spectra of the light unflavored mesons in the
framework of relativized quark model. By applying the auxiliary field method
to model, an analytical mass formula is proposed for the orbitally excited
unflavored mesons, by which a quasi-linear Regge-Chew-Frautschi plot is
predicted. We test the mass formula by fitting the observed data of the light
meson families', $\pi/b$,$\rho/a$,$\eta/h$ and $\omega/f$ in the ($L$,$M^{2}$)
planes, and find that the optimal values of the model parameters are
consistent with that of the relativized quark models in the literatures. An
agreement of the mass predicted by the mass formula with the experimental data
is achieved for the meson families considered.

It is also shown for large orbital quantum number $L$ that the trajectory
slope is inversely proportional to the confining parameter $a$, while the
intercept depends on the strong coupling $\alpha_{s}$, independent of
$a$.

The anomaly is observed in the fitted parameters when comforting the mass
formula with the experimental spectra of the $\eta/h$ trajectory. This may
imply that the unflavored states $\eta(548)0^{-+}$, $h_{1}(1170)1^{+-}$ and
others in the $\eta/h$ trajectory either have mixed with hidden $\bar{s}%
s$(or $\bar{c}c$) component in its internal structure or go beyond the
two-quark picture. 

\begin{description}
\item[\textbf{Acknoledgements}]
\end{description}

D. J thanks X. Liu for useful discussions. D. J is supported by the National Natural Science Foundation of China under the no. 11565023 and the Feitian Distinguished Professor Program of Gansu(2014-2016). C-Q. P is supported by the High-End Creative Talent Thousand People Plan of Qinghai Province, No. 0042801 and the Applied Basic Research Project of Qinghai Province, No. 2017-ZJ-748. A.H is supported by Grants-in-Aid for Scientific Research (Grants No. JP26400273(C) and JP25247036(A)).


\begin{thebibliography}{99}
\bibitem {ChewF:61}G. F. Chew, and S. C. Frautschi, {\it Phys. Rev. Lett.}
{\bf 7},394 (1961).
\bibitem {Collons:77}P. Collins, {\it An Introduction to Regge Theory and
High Energy Physics}, (Cambridge Univ. Press, Cambridge,1977,456).

\bibitem {AnisovichAS:PRD00}A. Anisovich, V. Anisovich, and A. Sarantsev,
{\it Phys.Rev. D} {\bf 62},051502 (2000);[hep-ph/0003113].

\bibitem {MasjuanAB:D12}Pere Masjuan,E. R. Arriola and W. Broniowski,
{\it Phys.Rev. D} {\bf 85}, 094006 (2012).

\bibitem {Sharov:13}G. Sharov, {\it String Models, Stability and Regge Trajectories
for Hadron States}, arXiv:1305.3985.

\bibitem {Selem:06}A. Selem and F. Wilczek, {\it Hadron systematics and emergent diquarks}, [hep-ph/0602128].

\bibitem {SonnenscheinW:JHEP14}J. Sonnenschein and D. Weissman, {\it JHEP}
{\bf 1408},013 (2014); [arXiv:1402.5603].

\bibitem {InopinS:2001}A. Inopin, and G. S. Sharov, {\it Phys. Rev. D} {\bf 63} ,054023 (2001).

\bibitem {TangN:PRD00}A. Tang and J. W. Norbury, {\it Phys.Rev. D} {\bf 62}
, 016006 (2000); [hep-ph/0004078].

\bibitem {GuoWW:D08}Xin-Heng Guo, Ke-Wei Wei, X. H. Wu, {\it Phys.Rev. D}
{\bf 78}, 056005 (2008).

\bibitem {JohnsonN:PRD79}K. Johnson and C. Nohl, {\it Phys.Rev. D} {\bf 79},
291 (1979).

\bibitem {BrisudovaBG:00} M.M. Brisudova, L. Burakovsky and T. Goldman, {\it
Phys.Rev. D} {\bf 61}, 054013 (2000).

\bibitem {Forkel:07a}H. Forkel, M. Beyer, and T. Frederico, {\it Int. J.
Mod. Phys. E} {\bf 16}, 2794 (2007).

\bibitem {Forkel:07b}H. Forkel, M. Beyer, and T. Frederico, {\it JHEP} {\bf 07},077 (2007).

\bibitem {KangS:75}J. S. Kang and H. J. Schnitzer, {\it Phys. Rev. D} {\bf 12},841 (1975).

\bibitem {MaungKN:D93}K. M. Maung, D. E. Kahana and J. W. Norbury,{\it Phys. Rev. D} {\bf 47}, 1182 (1993).

\bibitem {EbertFG:D09}D. Ebert, R. Faustov, and V. Galkin, {\it Phys. Rev. D}
{\bf 79},114029 (2009); [arXiv:0903.5183].

\bibitem {SilvestreSB:JPA2008}B. Silvestre-Brac, C. Semay and F. Buisseret, {\it J. Phys. A} {\bf 41}, 425301 (2008).

\bibitem {Afonin:A07}S. S. Afonin, {\it Mod. Phys. Lett. A} {\bf 22}, 1359 (2007).

\bibitem {Afonin:C07}S. S. Afonin, {\it Phys.Rev. C} {\bf 76}, 015202 (2007).

\bibitem {KlemptZ:P07}E. Klempt and A. Zaitsev, {\it Phys. Rep} {\bf 454}, 1 (2007).

\bibitem {ShifmanV:D08}M. Shifman and A. Vainshtein, {\it Phys. Rev. D} {\bf 77}, 034002(2008).

\bibitem {Afonin:JA07}S. S. Afonin, {\it Int. J. Mod. Phys. A} {\bf 22}, 4537 (2007).

\bibitem {Afonin:JA08}S. S. Afonin, {\it Int. J. Mod. Phys. A} {\bf 23}, 4205 (2008).

\bibitem {AfoninP14}S. S. Afonin and I. V. Pusenkov, {\it Phys. Rev. D} {\bf 90}, 094020 (2014).

\bibitem {AfoninPA14}S. S. Afonin and I. V. Pusenkov, {\it Mod. Phys. Lett. A} {\bf 29}, 1450193 (2014).

\bibitem {MasjuanAB14}P. Masjuan, E. R. Arriola and W. Broniowski, {\it EPJ Web Conf.} {\bf 73}, 04021 (2014).

\bibitem {Affonin:16}S. S. Afonin, {\it Acta Phys.Polon.Supp.} {\bf 9}, 597(2016).

\bibitem {Durand:D82}B. Durand and L. Durand, {\it Phys. Rev. D} {\bf 25}, 2312 (1982).

\bibitem {LichtenbergNP:pl82}D. B. Lichtenberg, W. Namgung, E. Predazzi, and
J. G. Wills, {\it Phys. Rev. Lett.} {\bf 48}, 1653 (1982).

\bibitem {Durand:D84}B. Durand and L. Durand, {\it Phys. Rev. D} {\bf 30}, 1904 (1984).

\bibitem {GodfreyIsgur:85}S. Godfrey and N. Isgur, {\it Phys. Rev. D} {\bf 32},189 (1985).

\bibitem {Jacobs:D86}S. Jacobs, M. G. Olsson, and C. J. Suchyta III, {\it Phys. Rev. D} {\bf 33}, 3338 (1986)

\bibitem {Bali01}G.S. Bali, {\it Phys. Rept.} {\bf 343}, 1 (2001).

\bibitem {KawanaiS:12}T. Kawanai and S. Sasaki, {\it Prog. Part. Nucl. Phys.}
{\bf 67}, 130 (2012).

\bibitem {AMPolyakov:87}A.M. Polyakov, {\it Gauge Fields and Strings},
(Harwood Academic Publishers,Poststrasse,1987).

\bibitem {GubankovaYD:PLB94}E. L. Gubankova and A. Yu. Dubin, {\it Phys. Lett. B} {\bf 334}, 180 (1994); [hep-ph/9408278].

\bibitem {MorgunovNS:PLB99}L. Morgunov, A. V. Nefediev and Yu. A. Simonov,
{\it Phys. Lett. B} {\bf 459}, 653 (1999).

\bibitem {SemaySN:2004}C. Semay, B. Silvestre-Brac and I. M. Narodetskii,{\it
Phys. Rev. D} {\bf 69},014003 (2004).

\bibitem {KalashnikovaN:PLB2000}Yu.S. Kalashnikova, A.V. Nefediev, {\it Phys.
Lett. B} {\bf 492}, 91 (2000).

\bibitem {Simonov:PL88} Yu. A. Simonov, Phys. Lett. 226, 151 (1988).
\bibitem {KalashnikovaN:PAN97} 42. Yu. S. Kalashnikova and A. V. Nefediev, Phys. At. Nucl. 60, 1389 (1997).

\bibitem {Goto:71}T. Goto, {\it Prog. Theor. Phys.} {\bf 46},1560 (1971).

\bibitem {GoddardGR:73}P. Goddard, J. Goldstone, C. Rebbi, C.B. Thorn, {\it Nucl.Phys. B} {\bf 56},109 (1973).

\bibitem {Nambu:74}Y. Nambu, {\it Phys. Rev. D} {\bf 10}, 4262 (1974).

\bibitem {Patrignani:2016} C. Patrignani {\it et al}. (Particle Data Group),{\it Chin.Phys. C} {\bf 40}, 100001 (2016).

\bibitem {VeseliO:B96}S. Veseli and M. G. Olsson, {\it Phys.Lett. B} {\bf 383}, 109 (1996).

\bibitem {FulcherCY:D93}L. P. Fulcher, Z. Chen and K. C. Yeong, {\it Phys. Rev. D} {\bf 47},4122 (1993).

\bibitem {HwangK:D96}D. S. Hwang and G.-H. Kim, {\it Phys. Rev. D} {\bf 53}, 3659 (1996).
\end{thebibliography}
\end{document}